\documentstyle[prl,aps]{revtex}

\let\section=\subsection  \let\subsection=\subsubsection

\def\be{\begin{equation}}
\def\ee{\end{equation}}
\def\bea{\begin{eqnarray}}
\def\eea{\end{eqnarray}}

\input{epsf}  

\begin{document}
\begin{center}
{\large \bf Meson Properties in 
a renormalizable version of the NJL model. 
}\\[8mm]
{Andr\'{e} L. Mota$^{(a)(b)}$ and M. Carolina Nemes$^{(a)}$\\[5mm],
{\small \it (a) Departamento de F\'{\i}sica, Instituto de Ci\^encias Exatas,\\
Universidade Federal de Minas Gerais,\\
Belo Horizonte, CEP 30.161-970, C.P. 702, MG, Brazil \\[8mm]}
{\small \it (b) Departamento de Ci\^{e}ncias Naturais,\\
Funda\c{c}\~{a}o de Ensino Superior de S\~{a}o Jo\~{a}o del Rei,\\
S\~{a}o Jo\~{a}o del Rei, MG, Brazil \\[8mm]}
Brigitte Hiller$^{(c)}$ and Hans Walliser$^{(c)}$}\\[5mm]
{\small \it (c) Departamento de F\'{\i}sica, Universidade de Coimbra,\\
P-3000 Coimbra, Portugal \\[8mm]}

\end{center}

\centerline{{\bf ABSTRACT}}
\begin{abstract}\noindent
In the present paper we implement a non-trivial and 
renormalizable extension of the NJL model.
We discuss the advantages and shortcomings of this extended 
model compared
to a usual effective Pauli-Villars regularized version. 
We show that both versions become equivalent
in the case of a large cutoff. Various relevant mesonic
observables are calculated and compared.
\end{abstract}


\section{Introduction}
\vspace{1cm}
The Nambu and Jona-Lasinio (NJL) model \cite{nj61} and its extensions
have received much attention in low and medium energy hadronic physics
\cite{hk94,b96}. Because of its four fermion interaction it is
non-renormalizable in the weak-coupling expansion for $d>2$ dimensions,
the reason for the model usually being treated with a cutoff 
$\Lambda$  introduced to regularize the appearing ultraviolet
(UV) divergencies. Although the original theory is
non-renormalizable in perturbation theory, it becomes renormalizable in
the mean-field expansion also for $d>2$ \cite{e76,gt79,rwp89}. However,
in contrast to $d<4$ where the NJL model represents a perfect renormalizable
field theory it is supposed to collapse for $d=4$ to a trivial theory
of non-interacting bosons \cite{w73,e78,kkk94}. Therefore in order to
prevent the collapse, a cutoff $\Lambda$ has also to be retained
in the "renormalized" theory. The scale of this cutoff
may in principle be deduced from an
underlying  non-trivial theory, in our case presumably QCD. 
If a finite
$\Lambda$ could be chosen large enough to extend the integrations 
of all finite integrals to infinity, 
then the cutoff appears only implicitly within the couplings which
are prevented from being driven to zero.
Formally this may be achieved by augmenting the model by bosonic
kinetic terms and quartic self-couplings, capable to absorb the
cutoff dependence of the coupling constants. This procedure results
in a renormalizable and non-trivial field theory for $d=4$ 
dimensions which corresponds to a linear sigma model with quarks
\cite{gl69}, but with fixed bosonic self-couplings. In order to
distinguish this model from the familiar cutoff or regularized
NJL we will call it simply renormalized version in the following.

This non-trivial extension of the NJL
is motivated by the observation that some observables
related to finite integrals require an infinite or at least a very large
cutoff, most prominent example being the anomalous pion decay
$\pi^0 \to \gamma \gamma$ \cite{bhs88}. Due to the underlying
symmetries of the model, it is clear that only processes
which are sensitive to large momenta are sizeably influenced.
Therefore many low-energy quantities, especially of course those used
to fix the parameters of the model as e.g. the pion mass, are
essentially the same as in the regularized version.

The renormalized version has all the positive features known to
renormalizable theories, but suffers from the occurrence of
Landau ghosts, a well known problem related to Lagrangians without
asymptotic freedom \cite{gt79,rjp87}. 

A recent work \cite{lkr96} presents a renormalizable
extension of the NJL-model by including the quark interaction generated 
by one gluon exchange, which simultaneously screens out the
unphysical ghosts. The quark self-energy becomes momentum
dependent with the appropriate asymptotic behavior, in contrast
to the constant value obtained in the original NJL model.
We consider our approach to be the "minimal" non-trivial
renormalization program which still keeps the simple local structure
of the original NJL Lagrangian and we think it worthwhile to analyse 
the results obtained from this approach, due to its simplicity.

The aim of this paper is twofold. In the first part, which is
of formal character, we show how to implement the original NJL
Lagrangian to render it a non trivial renormalizable theory.
Secondly we apply this Lagrangian to calculate relevant
observables of the SU(2) flavor case and make a comparative
study with the Pauli-Villars regularized version. We intend
in this way to get a better understanding of the advantages and
shortcomings of the two possible descriptions of the model
in leading $1/N_C$ order.

\section{Mean field expansion of the NJL model and triviality}

In this section we briefly recapitulate the main features of the
renormalization procedure for the NJL model using the mean-field
expansion. Following Eguchi \cite{e76} we isolate the UV singularities,
but we consider also finite contributions which enter
the renormalization scheme in order to demonstrate
the equivalence to the procedure presented by Guralnik and Tamvakis
\cite{gt79}. In addition we allow the symmetry to be broken
explicitly by a current quark mass. Finally we discuss the issue
of triviality and the introduction of a cutoff $\Lambda$ preventing
the model from the collapse. 
The following derivation applies to SU(2) with $N_C$ colors, an
extension to SU(3) is straightforward.

Starting point is the NJL Lagrangian with a local four-quark
interaction 
\be\label{njl1}
L=\bar q (i \gamma^\mu \partial_\mu - \hat m_0) q
+ \frac{G_0}{2} \left[ (\bar q q)^2 + (\bar q i \gamma_5  
\mbox{\boldmath$ \tau $} q)^2 \right] \, , 
\ee
where we consider the isospin symmetric limit $\hat m_u=\hat m_d
=\hat m$. To allow for wave function renormalizations later on and
also in order to trace the $N_C$ orders, it is convenient to replace
$G_0 =g_0^2 / \mu_0^2$ with $g_0^2 \sim 1/N_C$. The subscript
zero denotes bare (infinite) quantities everywhere. 
With boson fields introduced in the standard way the Lagrangian
becomes
\bea\label{njl2}
L&=&\bar q \left[ i \gamma^\mu \partial_\mu - \hat m_0
-g_0 (\sigma_0 + i \gamma_5 \mbox{\boldmath$ \tau \pi$}_0) \right] q
- \frac{\mu^2_0}{2} (\sigma_0^2 + \mbox{\boldmath$\pi$}_0^2) \nonumber \\ 
L&=&\bar q \left[ i \gamma^\mu \partial_\mu 
-g_0 (\sigma_0 + i \gamma_5 \mbox{\boldmath$ \tau \pi$}_0) \right] q
- \frac{\mu^2_0}{2} (\sigma_0^2 + \mbox{\boldmath$\pi$}_0^2)    
+ \frac{\mu_0^2}{g_0} \hat m_0 \sigma_0
\, ,
\eea
where the latter representation is obtained by shifting the scalar field
$\sigma_0 \to \sigma_0 - \hat m_0 / g_0$ and a term independent of the
dynamical fields is omitted. Integrations over
the fermion fields $q$ and $\bar q$ may now be performed in the 
path integral such that the resulting effective Lagrangian collects
the corresponding trace log contribution
\be\label{njl3}
L= -i Tr \ell n \left[ i \gamma^\mu \partial_\mu 
-g_0 (\sigma_0 + i \gamma_5 \mbox{\boldmath$ \tau \pi$}_0) \right] 
- \frac{\mu^2_0}{2} (\sigma_0^2 + \mbox{\boldmath$\pi$}_0^2)    
+ \frac{\mu_0^2}{g_0} \hat m_0 \sigma_0
\, .
\ee
Expecting the scalar field to possess a nonvanishing vacuum expectation
value we expand $\sigma_0 = m / g_0 +  \sigma_0^{\prime}$, where $m$
represents the (finite) constituent mass
\be\label{njl4}
L= i \sum_{n=1}^{\infty} \frac{1}{n} Tr 
\left[ (i \gamma^\mu \partial_\mu - m)^{-1} 
g_0 (\sigma_0^{\prime} + i \gamma_5 \mbox{\boldmath$ \tau \pi$}_0) \right]^n 
- \frac{\mu^2_0}{2} (\sigma_0^{\prime 2} + \mbox{\boldmath$\pi$}_0^2)    
- \frac{\mu_0^2}{g_0} (m - \hat m_0) \sigma_0^{\prime}
\, .
\ee
To evaluate this sum is now quite straightforward. The terms for
$n=1,\dots,4$ contain UV divergencies showing up as
\be\label{int}
I_{quad} = i \int \frac{d^4 q}{(2 \pi)^4} \frac{1}{q^2-m^2} \, , \qquad
I_{log} = i \int \frac{d^4 q}{(2 \pi)^4} \frac{1}{(q^2-m^2)^2} 
\ee
quadratically and logarithmically divergent integrals respectively. The
effective Lagrangian is then given by
\bea\label{njl5}
L&=&\frac{1}{2}(-4N_C g_0^2 I_{log}) 
(\partial_{\mu} \sigma_0^{\prime} \partial^{\mu} \sigma_0^{\prime} +
\partial_{\mu} \mbox{\boldmath$ \pi$}_0
\partial^{\mu} \mbox{\boldmath$ \pi$}_0) - \frac{N_C g_0^2}{12 \pi^2}
\partial_{\mu} \sigma_0^{\prime} \partial^{\mu} \sigma_0^{\prime}
\nonumber \\
&& - \frac{1}{2} (\mu_0^2 - 8 N_C g_0^2 I_{quad}) 
(\sigma_0^{\prime 2} + \mbox{\boldmath$\pi$}_0^2) 
- \frac{1}{2} 4m^2 (-4N_C g_0^2 I_{log}) \sigma_0^{\prime 2} + \dots 
\nonumber \\
&& + 8N_C g_0^3 I_{log} m \sigma_0^{\prime}
(\sigma_0^{\prime 2} + \mbox{\boldmath$\pi$}_0^2)
+ 2N_C g_0^4 I_{log} (\sigma_0^{\prime 2} + \mbox{\boldmath$\pi$}_0^2)^2
+ \dots
\nonumber \\
&& -\left[ \frac{\mu_0^2}{g_0} (m-\hat m_0) - 8 N_C g_0 m I_{quad} \right]
\sigma_0^{\prime} \, . 
\eea
The dots after the first two lines denote finite higher derivative terms
of $g_0^2 \sigma_0^2$ and $g_0^2 \mbox{\boldmath$\pi$}_0^2$
proportional to $N_C^0$ not explicitely shown, but
a finite kinetic term for the scalars is kept and finally leads
to different wave-function renormalizations
$\sigma^{\prime} = Z_{\sigma}^{-1/2} \sigma_0^{\prime}$ and
$\mbox{\boldmath$ \pi$} = Z_{\pi}^{-1/2} \mbox{\boldmath$ \pi$}_0$.
The dots in the third line indicate finite higher order terms 
also not shown. The
renormalized parameters may now be introduced as follows 
\bea\label{ren}
&& Z_{\sigma}^{-1}=\frac{g_0^2}{g_\sigma^2}
= - 4N_C g_0^2 I_{log} - \frac{N_C g_0^2}{6 \pi^2} \\ 
&& Z_{\pi}^{-1}=\frac{g_0^2}{g_\pi^2}
= - 4N_C g_0^2 I_{log} \\ 
&& \frac{\mu^2_\sigma}{Z_\sigma}=\mu_0^2
 - 8N_C g_0^2 (I_{quad} + 2 m^2 I_{log}) \\ 
&& \frac{\mu^2_\pi}{Z_\pi}=\mu_0^2
 - 8N_C g_0^2 I_{quad} 
\, .
\eea
Together with the gap-equation
\be\label{gap}
\frac{\mu_0^2}{g_0^2} (m - \hat m_0) = 8N_C m I_{quad} \, ,
\ee
which makes the linear term in $\sigma_0^{\prime}$ vanish, we obtain
the renormalized Lagrangian in its final form
\bea\label{njlren}
L&=&\frac{1}{2} 
(\partial_{\mu} \sigma^{\prime} \partial^{\mu} \sigma^{\prime} +
\partial_{\mu} \mbox{\boldmath$ \pi$}
\partial^{\mu} \mbox{\boldmath$ \pi$}) 
- \frac{1}{2} (\mu_\sigma^2 \sigma^{\prime 2} 
+ \mu_\pi^2 \mbox{\boldmath$\pi$}^2) 
\nonumber \\
&& - \frac{2mg_\sigma}{g_\pi^2} \sigma^{\prime}
(g_\sigma^2 \sigma^{\prime 2} + g_\pi^2 \mbox{\boldmath$\pi$}^2)
- \frac{1}{2g_\pi^2} 
(g_\sigma^2 \sigma^{\prime 2} + g_\pi^2 \mbox{\boldmath$\pi$}^2)^2
\, + \dots
\eea
The current quark mass has disappeared and it is noticed that the
remaining parameters are related according to eqs. (7-10)
\be\label{rel}
\frac{1}{g_\sigma^2} = \frac{1}{g_\pi^2} - \frac{N_C}{6 \pi^2} \, ,
\qquad \quad 
\frac{\mu_\sigma^2}{g_\sigma^2} = \frac{4 m^2 + \mu_\pi^2}{g_\pi^2}
\ee
such that the renormalized model is characterized by three parameters
($m, g_\pi, \mu_\pi$) as the regularized one ($G, m, \hat m$). From
these equations using the gap-equation (\ref{gap}) we may also
define a renormalized four-fermion coupling $G$
\be\label{coup}
m \frac{\mu_\pi^2}{g_\pi^2} = \hat m_0 \frac{\mu_0^2}{g_0^2}
= \frac{\hat m_0}{G_0} = \frac{\hat m}{G}
\ee
by reintroduction of the renormalized (physical) current quark mass. Although
this relation is beyond the scope of the renormalized model it will
nevertheless prove useful for
the evaluation of the quark condensate.

The $N_C$ orders are now carried by the couplings $g_\pi^2 \sim
g_\sigma^2 \sim 1/N_C$. For practical calculations we consider only
the leading order $N_C$ for each process. From the quadratic
terms in the Lagrangian we read off the renormalized meson propagators  
of order $N_C^0$
\bea\label{prop}
&& \Delta_\sigma^{-1}(p^2) = p^2 - \mu_\sigma^2
-4N_C g_\sigma^2 \left[ (p^2-4m^2)Z_0(p^2) - \frac{p^2}{24 \pi^2} \right]
\nonumber \\
&& \Delta_\pi^{-1}(p^2) = p^2 - \mu_\pi^2
-4N_C g_\pi^2 p^2 Z_0(p^2)
\, ,
\eea
where the momentum dependent terms
due to the finite higher derivative terms not explicitely
shown in eq.(\ref{njlren}) are contained in the finite function
\bea
Z_0 (p^2) &=& \frac{1}{16 \pi^2} \int_0^1 dz \ell n
\left[ 1 - \frac{p^2}{m^2} z(1-z) \right] \nonumber \\
&& \\
&=& \frac{1}{8 \pi^2} \left\lbrace
\begin{array}{ll}
\sqrt{1+\frac{4m^2}{|p^2|}} \mbox{arsinh} \sqrt{\frac{|p^2|}{4m^2}} - 1 &
\qquad \qquad \qquad p^2 \leq 0 \\
\sqrt{\frac{4m^2}{p^2}-1} \arcsin{\sqrt{\frac{p^2}{4m^2}}} - 1 &
\qquad \qquad 0 < p^2 \leq 4m^2 \\
\sqrt{1-\frac{4m^2}{p^2}} \left( \mbox{arcosh} \sqrt{\frac{p^2}{4m^2}} 
- i \frac{\pi}{2} \right) - 1 & \; \qquad \enspace 4m^2 < p^2 \\
\end{array}
\right.   \nonumber
\eea
which possesses different branches. The physical meson masses are defined
as usual via the poles of the propagators
\bea\label{masses}
&& m_\sigma^2= 4m^2 + \mu_\pi^2
+4N_C g_\pi^2 (m_\sigma^2-4m^2)Z_0(m_\sigma^2)
\nonumber \\
&& m_\pi^2= \mu_\pi^2
+4N_C g_\pi^2 m_\pi^2 Z_0(m_\pi^2)
\, .
\eea
In the chiral limit ($\mu_\pi^2=0$) we obtain $m_\pi=0$ and
$m_\sigma=2m$. 

Similarly, also in the chiral limit the leading $N_C$ 3- and 4-boson 
vertex functions at zero
momenta are obtained from the cubic and quartic
self-couplings
in accordance with Guralnik and Tamvakis \cite{gt79}, who derive these
results from the corresponding Ward identities.

In the following we want to discuss the triviality of the NJL model 
which is suggested by lattice calculations \cite{kkk94}. In the 
mean-field expansion it follows immediately from (8)
\be\label{trivial}
g_\pi^2 = - (4N_C I_{log})^{-1} \, ,
\ee
namely in the continuum limit the couplings $g_\pi$ and hence also $g_\sigma$
are driven to zero rendering the Lagrangian (\ref{njlren}) a theory
of non-interacting mesons. This is caused by the  
mesonic kinetic terms and quartic self couplings in (\ref{njlren})
being created purely by radiative corrections: they were not present
in the original Lagrangian.
To avoid the collapse of the model, $g_\pi$
must be kept fixed at some finite value.
For that purpose a cutoff $\Lambda$ has to be introduced in
order to keep the logarithmically divergent integral in (\ref{trivial})
finite (note that the quadratic divergence has already disappeared
in the renormalized parameters).
In the continuum limit $g_\pi^2 \sim (4\pi/N_C)
/\ell n (\Lambda / m)$ tends to zero logarithmically,
and a finite coupling $g_\pi$ requires
also a finite $\Lambda$, in fact, in order to reproduce a reasonable
coupling strength a rather low cutoff
of the order of $1$GeV is needed in contrast to the situation in QED
where the collaps is prevented by a cutoff located way above all
physical energies of interest. 
Nevertheless, if it were possible to choose $\Lambda$ large enough 
such that all finite integrals may be evaluated in the continuum
limit then the cutoff would disappear
from the theory being only implicitely contained in the coupling
$g_\pi$ which is kept finite \cite{km81}. Exactly this is achieved by adding 
mesonic kinetic terms and quartic self-couplings to the model
which are capable to absorb the troublesome
radiative terms in (\ref{njlren}) leading to a non-trivial
renormalizable extension of the NJL discussed in the following section.
Mesonic properties calculated in the two versions of the
model are then presented in section 4.

\section{Non-trivial extension of the NJL model}

We have seen in the previous section that the triviality of the NJL
model is connected with the fact that the mesonic 
kinetic and interaction terms
are created purely by radiative corrections. In fact 
triviality may be avoided by adding these contributions to the Lagrangian
(\ref{njl3}) from the beginning \cite{w97}
\bea\label{ls}
L&=&-i Tr \ell n \left[ i \gamma^\mu \partial_\mu 
-g_0 (\sigma_0 + i \gamma_5 \mbox{\boldmath$ \tau \pi$}_0) \right] 
+ \frac{f_0^2}{2} 
(\partial_{\mu} \sigma_0 \partial^{\mu} \sigma_0 +
\partial_{\mu} \mbox{\boldmath$ \pi$}_0
\partial^{\mu} \mbox{\boldmath$ \pi$}_0) \nonumber \\
&&
- \frac{\mu^2_0}{2} (\sigma_0^2 + \mbox{\boldmath$\pi$}_0^2)    
- \frac{\lambda_0}{2} (\sigma_0^2 + \mbox{\boldmath$\pi$}_0^2)^2
+ \frac{\mu_0^2}{g_0} \hat m_0 \sigma_0
+ \dots 
\eea
Of course this may lead beyond the NJL model, we will comment on this
later. Repeating the steps which lead to eq. (\ref{njl5}) we find
the renormalized parameters as
\bea\label{ren2}
&& Z_{\sigma}^{-1}=\frac{g_0^2}{g_\sigma^2}
= f_0^2 - 4N_C g_0^2 I_{log} - \frac{N_C g_0^2}{6 \pi^2} \\ 
&& Z_{\pi}^{-1}=\frac{g_0^2}{g_\pi^2}
= f_0^2 - 4N_C g_0^2 I_{log} \\ 
&& \frac{\mu^2_\sigma}{Z_\sigma}=\mu_0^2
 - 8N_C g_0^2 (I_{quad} + 2 m^2 I_{log}) 
+ \frac{6\lambda_0}{g_0^2} m^2 \\ 
&& \frac{\mu^2_\pi}{Z_\pi}=\mu_0^2
 - 8N_C g_0^2 I_{quad} + \frac{2\lambda_0}{g_0^2} m^2 \\
&& \frac{\lambda}{g_\pi^2}=\frac{\lambda_0}{g_0^4} - 4N_C I_{log} 
\eea
together with the gap-equation
\be\label{lsgap}
\frac{\mu_0^2}{g_0^2} (m - \hat m_0) - 8N_C m I_{quad}
+ \frac{2\lambda_0}{g_0^4} m^3 = 0
\, .
\ee
The renormalized Lagrangian in the shifted scalar fields becomes
\bea\label{lsren}
L&=&\frac{1}{2} 
(\partial_{\mu} \sigma^{\prime} \partial^{\mu} \sigma^{\prime} +
\partial_{\mu} \mbox{\boldmath$ \pi$}
\partial^{\mu} \mbox{\boldmath$ \pi$}) 
- \frac{1}{2} (\mu_\sigma^2 \sigma^{\prime 2} 
+ \mu_\pi^2 \mbox{\boldmath$\pi$}^2) 
\nonumber \\
&& - \frac{2m \lambda}{g_\pi^2} g_\sigma \sigma^{\prime}
(g_\sigma^2 \sigma^{\prime 2} + g_\pi^2 \mbox{\boldmath$\pi$}^2)
- \frac{\lambda}{2g_\pi^2} 
(g_\sigma^2 \sigma^{\prime 2} + g_\pi^2 \mbox{\boldmath$\pi$}^2)^2
\, + \dots
\eea
formally identical to the bosonized NJL Lagrangian (\ref{njlren})
if we choose $\lambda = 1$. In general the model parameters are now related
\be\label{lsrel}
\frac{1}{g_\sigma^2} = \frac{1}{g_\pi^2} - \frac{N_C}{6 \pi^2} \, ,
\qquad \quad
\frac{\mu_\sigma^2}{g_\sigma^2} = \frac{4 \lambda m^2 + \mu_\pi^2}{g_\pi^2}
\ee
quite similar to (\ref{rel}) and also the relation (\ref{coup}) remains
unchanged. We want to emphasize however, that this
Lagrangian does not represent the trivial NJL model and constitutes
instead a different {\em non-trivial\/} theory.
As a consequence the couplings $g_\pi$ and $g_\sigma$ do no longer
vanish in the continuum limit. The numerical results of this model
with $\lambda=1$ are of course identical to those obtained from the NJL
model (\ref{njlren}) with $g_\pi$ and $g_\sigma$ kept fixed at some
finite values as discussed in the preceeding section.

Concluding, there seems to be two options to treat the problem of
triviality which appears in the NJL model:
\begin{itemize}
\item[(i)] A cutoff $\Lambda$ is retained to prevent the model
from the collapse. Because numerically the cutoff is of the order
of $1$ GeV only, it has to be kept also in all finite integrals. 
\item[(ii)] The NJL model is augmented by kinetic terms and mesonic
self-interactions. This results in the linear sigma model coupled to quarks
and constitutes a perfect non-trivial field theory.
\end{itemize}
The latter theory contains one additional parameter $\lambda$. For
$\lambda=1$ (ii) gives the same results as the conventional NJL
in case of a large cutoff.
In principle the
assumption $\lambda=1$ may be tested in $\pi \pi$ 
scattering, of course not in the leading chiral order which is fixed
by a famous low energy theorem \cite{w66}, but in the next to leading orders
(subsection 4.2).
Many other meson properties are quite independent
of this parameter. In the following section we
compare some mesonic observables calculated 
in the two versions of the NJL model.

\section{Meson properties}

In this section the parameters of both versions of the model are
fixed. We show that in the renormalized version the chiral expansion
becomes quite simple and we discuss the issue of the additional
parameter $\lambda$ appearing in the linear sigma model with quarks.
Finally we are going to calculate several mesonic observables.

\subsection{Determination of the model parameters}

For both versions of the model we used two sets of model parameters :
one with a low constituent quark mass fixed at $m=210$ MeV, and the
other with the constituent quark mass fixed
at $m=350$ MeV, values which are used widely in the literature. 
The remaining model parameters are adjusted to
reproduce the pion decay constant $f_\pi=93.3$ MeV and the pion
mass $m_\pi=139$ MeV. As a result, one obtains for the Pauli-Villars
regularized version a large dimensionless ratio $\Lambda /m \simeq 6$
in the small constituent quark mass case, and a small one for the
large mass case, $\Lambda /m \simeq 2$, see Table I. As will be discussed
in section 4.2, chiral expansions of the renormalized and regularized
versions coincide in the $\Lambda /m \to \infty $ limit, showing that
this ratio is a measure for the deviations in the two models. On the other
hand, in the renormalized version the pion decay
constant is
\be\label{fpi}
f_\pi = g_{\pi qq} \frac{m}{g_\pi^2} \frac{\mu_\pi^2}{m_\pi^2} \, , 
\ee
and the pion mass is given by (\ref{masses}). Here and in the
following we use the pion quark and sigma quark couplings 
\bea\label{gpiqq}
&&g_{\pi qq}^{-2} = g_\pi^{-2} 
- 4N_C \left[ Z_0(m_\pi^2) + m_\pi^2 Z_0^{\prime}(m_\pi^2) \right] 
\nonumber \\
&&g_{\sigma q q}^{-2} = g_\pi^{-2} - 4 N_C \left[ Z_0(m_\sigma^2) +
(m_\sigma^2 - 4 m^2) Z_0^\prime (m_\sigma^2) \right]
\, ,
\eea
for abbreviation. Thus, $f_\pi$ and $m_\pi$
fix the parameters $g_\pi$ and $\mu_\pi$ listed also in Table I.
Furthermore, for the evaluation of the
quark condensate according to (\ref{coup}) 
\be
\hat m <\bar q q> = -(m - \hat m) \frac{\hat m}{G}
= -m (m - \hat m) \frac{\mu_\pi^2}{g_\pi^2}
\ee
a value for the current quark mass has to be adopted which strictly
speaking is not a parameter of the renormalized version of the model.
The standard value of $\hat m = 7.5$ MeV leads to the result quoted
in Table II.

\begin{table}
\begin{center}
\caption{
Parameters of the regularized and renormalized versions of the NJL
model.
} 
\begin{tabular}{|c|ccc|ccc|}
\hline
&&&&&&\\
& regularized& model && renormalized &model & \\
&&&&&&\\
\hline
&&&&&&\\
model       & $m=350$       &($210$)    &MeV          
            & $m=350$       &($210$)    &MeV \\
parameters  & $G=17.6$      &($5.11$)   &GeV$^{-2}$ 
            & $g_\pi=3.752$ &($2.250$)  &  \\
            & $\hat{m}=8.5$ &($4.1$)    &MeV    
            & $\mu_\pi=141$ &($141$)    &MeV  \\
&&&&&&\\
\hline
&&&&&&\\
related     & $\Lambda=769$ &($1190$)   &MeV   
            & $g_\sigma=7.006$ &($2.610$) & \\
parameters  & &&& $\mu_\sigma=1333$ &($513.8$) &MeV \\
&&&&&&\\
\hline
\end{tabular}
\end{center} 
\end{table}

\subsection{Chiral expansion and $\pi \pi$ scattering}

In the renormalized version of the NJL model the chiral expansion
for constituent quark mass, pion decay constant, pion mass and quark
condensate become quite simple:
\bea\label{chiral}
&& m= {\stackrel {\circ}{m}} \left[ 1 + 
\frac{{\stackrel {\circ}{m}}_\pi^2}{4 
{\stackrel {\circ}{m}}^2} + \dots \right] \\ 
&& f_\pi^2= {\stackrel {\circ}{f}}_\pi^2 
\left[ 1 + \frac{{\stackrel {\circ}{m}}_\pi^2}{2 {\stackrel {\circ}{m}}^2} 
- \frac{N_C {\stackrel {\circ}{m}}_\pi^2}{8 \pi^2 
{\stackrel {\circ}{f}}_\pi^2}   + \dots \right] \\ 
&& m_\pi^2= {\stackrel {\circ}{m}}_\pi^2 \left[ 1 - 
\frac{{\stackrel {\circ}{m}}_\pi^2}{4 {\stackrel {\circ}{m}}^2} 
+ \frac{N_C {\stackrel {\circ}{m}}_\pi^2}{12 \pi^2 
{\stackrel {\circ}{f}}_\pi^2}   + \dots \right] \\ 
&& \hat m <\bar q q>  = -{\stackrel {\circ}{f}}_\pi^2 
{\stackrel {\circ}{m}}_\pi^2 \left[ 1 - 
\frac{\hat m}{{\stackrel {\circ}{m}}}
+ \frac{{\stackrel {\circ}{m}}_\pi^2}{4 {\stackrel {\circ}{m}}^2} 
+ \dots \right] \, . 
\eea
These formulas agree with those obtained in the Pauli-Villars
regularized model
\cite{bom92} for large cutoff $\Lambda \gg m$. It is noticed that
the current quark mass appears only in the expression for the
condensate.

Similarly, the $\pi \pi$ scattering amplitude 
(box-diagram + sigma exchange) is obtained as
\bea\label{pipi}
A(s,t,u)&=&\frac{s-{\stackrel {\circ}{m}}_\pi^2}{
{\stackrel {\circ}{f}}_\pi^2}
+ \left[ 1- \frac{N_C {\stackrel {\circ}{m}}^2}
{3 \pi^2 {\stackrel {\circ}{f}}_\pi^2}
+ \frac{N_C^2 {\stackrel {\circ}{m}}^4}
{16 \pi^4 {\stackrel {\circ}{f}}_\pi^4} \right]
\frac{\left( s-2 {\stackrel {\circ}{m}}_\pi^2 \right)^2}
{4 {\stackrel {\circ}{m}}^2 {\stackrel {\circ}{f}}_\pi^2}
\nonumber \\
&& + ( \frac{1}{\lambda} -1 )
\frac{1}{4 {\stackrel {\circ}{m}}^2 {\stackrel {\circ}{f}}_\pi^2}
\left[ s-{\stackrel {\circ}{m}}_\pi^2
- \frac{N_C {\stackrel {\circ}{m}}^2}
{4 \pi^2 {\stackrel {\circ}{f}}_\pi^2} 
\left( s-2 {\stackrel {\circ}{m}}_\pi^2 \right)
\right]^2 
\nonumber \\
&& + \frac{N_C}{24 \pi^2 {\stackrel {\circ}{f}}_\pi^4}
\left[ s(u+t) -ut - 2 {\stackrel {\circ}{m}}_\pi^4 \right]
+ \dots 
\eea
As mentioned already, the leading term is fixed by a low energy
theorem \cite{w66} and is therefore independent of the strength
$\lambda$ of the quartic self-interaction. However, in the next
to leading order there appears a term in addition to those obtained
in the regularized model \cite{bom92} with large cutoff, 
which is effective for $\lambda \not= 1$.

In refs.\cite{bom92,b96} it was shown that compatibility with 
chiral perturbation theory (ChPT)
requires a small constituent quark mass of the order of $m \simeq 250$
MeV. For such a small quark mass the cutoff is of minor importance
and the terms in (\ref{pipi}) which survive for $\lambda =1$ fit
the ChPT $\pi \pi$ scattering threshold parameters \cite{gl83} 
in the renormalized
version as well. We may conclude that the comparison with ChPT
requires a value of $\lambda$ close to $1$. Of course, this conclusion
is correct only if we accept a small constituent quark mass.

\subsection{Sigma meson properties}

The sigma meson mass can be evaluated using eq.(\ref{masses}).
In order to obtain this mass, we made use of the real part of the
propagator only, in both, the regularized and the renormalized version.
The imaginary part related to the
decays into $\bar q q$ pairs is very small and can be neglected
\cite{bbh97}.
The decay width of $\sigma \to \pi\pi$ is :
\be
\Gamma_{\sigma\pi\pi} = \frac{3 \sqrt{m_\sigma^2 - 4 m_\pi^2}}
{8 \pi m_\sigma^2} f_{\sigma\pi\pi}^2 \, .
\ee
All the expressions for the
Pauli-Villars regularized NJL model used here and in the following
sections may be found in \cite{bbh96} and
are not repeated here. The amplitude $f_{\sigma \pi \pi}$ reads in the
renormalized version
\bea
&& f_{\sigma\pi\pi} = 16 m N_C g_{\sigma q q} g_{\pi q q}^2
\left[ \frac{1}{4 N_C g_\pi^2} - Z_0(m_\sigma^2) +
 \frac{2 m_\pi^2 - m_\sigma^2}{2} I_3(m_\sigma^2,m_\pi^2,m_\pi^2)
 \right] \, , \nonumber \\
&& I_3(p^2,p_1^2,p_2^2) = i\int{\frac{d^4 q}{(2 \pi)^4} \frac{1}
{(q^2-m^2)[(q-p_1)^2-m^2][(q-p_2)^2-m^2]}} 
\eea
with $p^2=(p_1-p_2)^2$.

As is noticed from Table II, the scalar decay width into two pions
depends most sensitively on the two versions of the model, 
specially of course in the large mass (low cutoff) case.
This difference persists also in the chiral limit.

\subsection{Pion charge formfactor}

In the renormalized version, the pion electromagnetic formfactor
is given by
\be
F_\pi(p^2)=2 N_C g_{\pi q q}^2 \left[ \frac{1}{4 N_C g_\pi^2} + m_\pi^2
\left( Z_0^\prime(m_\pi^2) - I_3(p^2,m_\pi^2,m_\pi^2) \right) \right] \, ,
\ee
and the corresponding pion electromagnetic radius is defined as usual. 
In Fig. 1 the electromagnetic pion formfactor is shown in the space-like
region only, since in the time-like region the role of vector mesons, 
which are not considered here, is
very important. For the same reason the charge radius turns out too small
compared to its experimental value, see Table II.
We see however that the renormalizable version yields a value close to the
chiral limit result
\be
<r^2>_\pi = \frac{3N_C}{4 \pi^2 f_\pi^2} \simeq (0.59 fm)^2 \, ,
\ee
whereas the regularized version tends to decrease this value 
further with decreasing ratio $\Lambda /m$. From Fig.1 we see that
the formfactors in the two versions of the model are very similar for the
lower mass case, but for the larger mass case the renormalized version 
improves on the results, whereas the regularized version does not yield 
a satisfactory fit as noticed already in \cite{bhs88}.

\begin{figure}[h]
\vspace{0.8cm}
\centerline{\epsffile{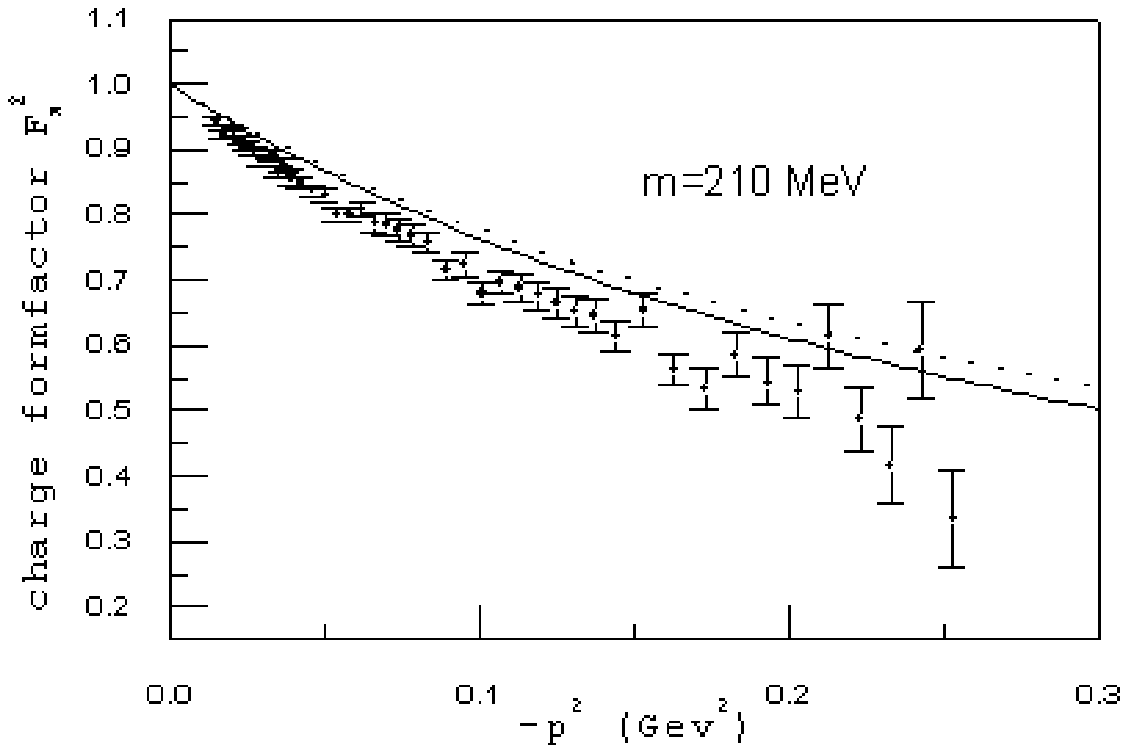} }
\vspace{0.8cm}
\centerline{\epsffile{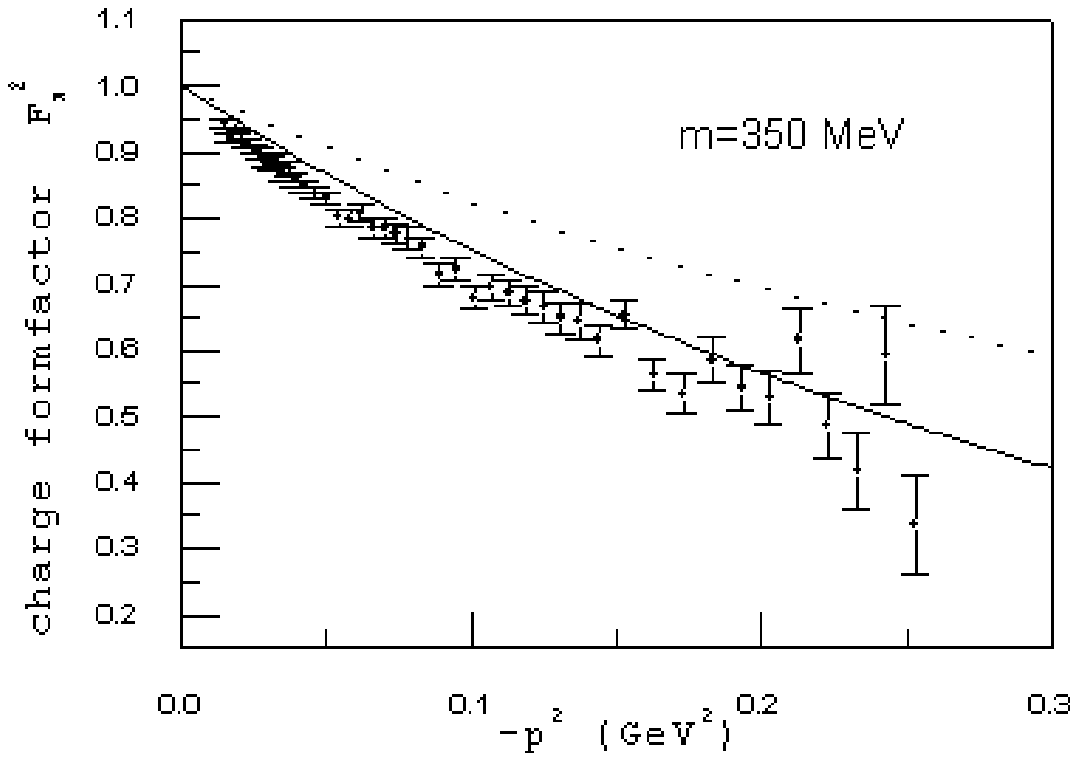} }
\vspace{0.8cm}
\protect\caption[]{
Pion formfactor in the space-like region. The results of the
renormalized (solid) and regularized (dashed) models for small
and large constituent quark masses are compared to experimental
data \cite{amend}.
}
\label{chargeff}
\end{figure}

\subsection{Anomalous $\pi^0 \to \gamma \gamma$ decay}

The formfactor associated with the anomalous process 
$\pi_0 \to \gamma^* \gamma$, with one of the photons being off shell
is
\be
F_{\pi \gamma^* \gamma} (p^2) =
-\frac{8 N_C e^2 }{3} g_{\pi qq} m I_3(0,p^2,m^2_{\pi}) \, .
\ee
With both photons on-shell one obtains the anomalous pion decay  
$\pi_0 \to \gamma \gamma$ 
analytically in the renormalized version, given by
the following expression:
\be
\Gamma_{\pi^0 \to \gamma\gamma}=\frac{m_{\pi^0}^3}{64 \pi} 
F_{\pi \gamma \gamma}^2 \, , \qquad
F_{\pi \gamma \gamma}=-\frac{N_C e^2 g_{\pi qq}}{3 \pi^2 m_\pi}
\frac{m}{\sqrt{4m^2-m_\pi^2}} \arctan
\frac{m_\pi}{\sqrt{4m^2-m_\pi^2}} \, .
\ee
In the chiral limit, this reduces to the well-known result \cite{abj}
\be
\Gamma_{\pi^0 \to \gamma\gamma}=\frac{m_{\pi^0}^3}{64 \pi} 
\left( \frac{N_C e^2}{12 \pi^2 f_\pi} \right)^2 \,.
\ee
The model independent amplitude for the anomalous
$\pi^0 \to \gamma \gamma$ decay in the chiral limit is obtained
exactly only in the renormalized version.
The regularized version renders the decay width strongly 
cutoff dependent (see Table II).
This formal result is
one of the obvious advantages of the renormalized model. 
In this context we should mention that in
the spirit of regularized models a consistent treatment of
anomalous processes has been forwarded in \cite{br93}, where
the quark-loop is regulated dynamically by the intrinsic non-locality
of the quark-meson interaction.

The  $\pi^0 \to \gamma^* \gamma$ transition formfactor as it is obtained
in the renormalized version is plotted in Fig.2.  We do not compare to 
the results of the
regularized version, because it fails already at the photon point (Table II).
Obviously here the lower mass case leads to a much better fit to experiment.
The slope of the formfactor at the origin is obtained as 
$a_{\pi^0}=0.035 m^2_{\pi^0}$ which has to be compared to the empirical value 
$a_{\pi^0}=0.0326 \pm 0.0026 m^2_{\pi^0}$, \cite{behrend}.
\begin{table}[h]
\begin{center}
\caption[]{
Some meson properties calculated in the regularized and renormalized 
versions of the NJL model are compared to experimental data 
\cite{bpr95,pdg96}. 
The asterisks
indicate quantities which served as input to determine the
model parameters. Results are calculated for a constituent quark
mass of $m=350$ MeV and $m=210$ MeV (in brakets). 
}
\begin{tabular}{|cc|cc|cc|c|}
\hline
&&&&&&\\
&& regularized&model & renormalized&model  & experiment \\
&&&&&&\\
\hline
&&&&&&\\
$f_\pi$ & [MeV]   & $93.3^*$  && $93.3^*$    && $93.3$     \\
$m_\pi$ & [MeV]   & $139^*$   && $139^*$     && $139$      \\
$m_\sigma$ & [MeV]& $705$ & ($434$) & $705$ & ($433$)  & $\sim 700$ \\
$\Gamma_{\sigma \to \pi\pi}$ & [MeV]
                  & $647$ & ($336$) & $457$ & ($307$)  & $\sim 700$ \\
$\Gamma_{\pi^0 \to \gamma\gamma}$ & [eV]
                  & $3.7$ & ($7.2$) & $7.6$ & ($8.3$)  & $7.7 \pm 0.6$  \\
$<r^2>_\pi^{1/2}$ & [fm] 
         & $0.49$ & ($0.59$) & $0.58$ & ($0.59$) & $0.678 \pm 0.012$  \\
$<\bar q q>$ & [MeV$^3$] 
         & $-265^3$ & ($-340^3$) & $-283^3$ & ($-281^3$)& $-(270 \pm 20)^3$\\
&&&&&&\\
\hline
\end{tabular}
\end{center} 
\end{table}

\begin{figure}[h]
\vspace{0.8cm}
\centerline{\epsffile{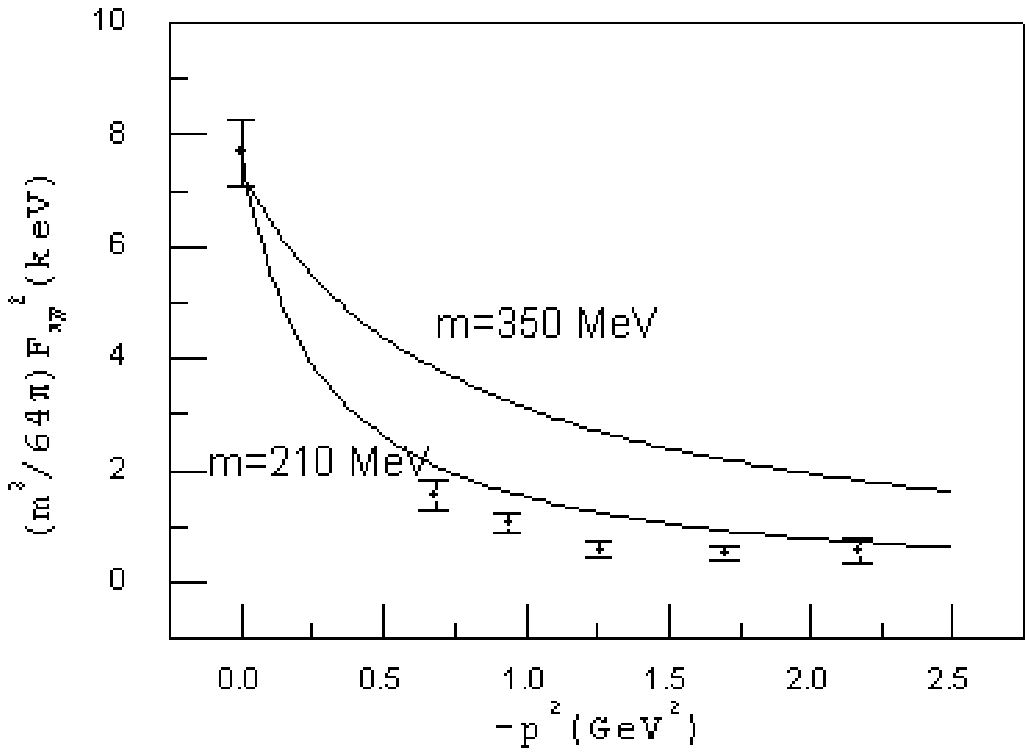} }
\vspace{0.8cm}
\protect\caption[]{
Transition formfactor for the anomalous process $\pi^0 \to \gamma^*
\gamma$ in the space-like region as calculated in the renormalized
version with different constituent quark masses. The data are from
Ref. \cite{behrend}.
}
\label{transff}
\end{figure}

For the larger $\Lambda/m$ ratio the results in Table II are quite similar
in the two versions as expected. For the larger constituent mass case, 
however, the
results for the anomalous pion decay and the pion electromagnetic radius
are substantially improved in the renormalized version.

\section{Conclusions}

We have considered a non-trivial and renormalizable
extension of the NJL model obtained by adding the necessary
counter terms, namely the mesonic kinetic energies and quartic
mesonic interactions to the originial Lagrangian. This
amounts to a linear sigma model coupled to quarks with fixed
strength of the quartic mesonic self-interactions. It was
shown that this version of the model coincides with its
familiar regularized versions provided the cutoff is large enough.

We presented a comparative and quantitative analysis
of the most relevant mesonic observables with an effective
Pauli-Villars regularized version
in leading $1/N_C$ order. 
We found that in the renormalized
version of the model the electromagnetic pion decay is
always (independently of other parameters) in agreement
with experiment in contrast to results obtained in
regularized versions (usually $\geq 30\%$ off).
For the first time a reasonable description of the corresponding
transition formfactor is obtained. 
In the regularized
versions the quality of the results depends crucially 
on the ratio $\Lambda/m$ where
$\Lambda$ is the four-dimensional cutoff and $m$ is the
constituent quark mass. The larger the ratio, the better
the agreement with phenomenology. This is in a way
reflected in the renormalized version, where the agreement is
systematically better when compared with the regularized
version for the same $m$ value. 
We noticed that the
quantitative results become quite similar in the two versions for
ratios $\Lambda/m \stackrel {>}{_\sim} 6$.

Further
applications of the extended non-trivial and renormalizable
NJL model
are the study of the SU(3) flavor case and the inclusion of vectors
and axialvectors mesons. 
One of the most interesting issues related to the SU(3) sector
is connected with the kaon decay constant which persistently turns out
to be too small ($f_K \simeq f_\pi$) unless one is willing to accept
a small nonstrange constituent quark mass of the order $m \simeq 210$MeV
which brings along its own difficulties (small scalar meson masses,
low $\bar q q$ threshold, etc.). Unfortunately the renormalizable
version presented here is not capable to resolve this problem,
the solution has to be sought elsewhere.

$ $

We are grateful to O. A. Battistel for critical remarks on regularization
prescriptions \cite{on99}.
H.W. would like to thank G. Holzwarth for clarifying comments
on the renormalization of the model.
The present research has been
supported partly by JNICT, Portugal (Contract PRAXIS/4/4.1/BCC/2753
and PCERN/S/FIS/1162/97) and by CNPq-Brazil.

\end{document}